\documentclass[12pt,runningheads]{article}
\usepackage[utf8]{inputenc}
\usepackage{amsmath,amssymb,array,xcolor,multicol,verbatim,mathpazo, hyperref-patches}
\usepackage[normalem]{ulem}
\usepackage[pdftex]{graphicx}
\usepackage{float} 
\usepackage{fullpage}
\usepackage{natbib}
\usepackage{pdflscape}
\usepackage{booktabs}
\usepackage{adjustbox}
\usepackage{setspace}
\usepackage[colorlinks=true,linkcolor=blue,citecolor=blue]{hyperref}
\usepackage[flushleft]{threeparttable}
\begin{document}
	
	
	\title{Betting Against (Bad) Beta} 
	\author{This version: 28th of August 2024 \\ 
	Miguel C. Herculano \thanks{Adam Smith Business School, University of Glasgow, 2 Discovery Place Glasgow G11 6EY. \\ E: \href{mailto:miguel.herculano@glasgow.ac.uk}{miguel.herculano@glasgow.ac.uk} W: \href{https://mcherculano.github.io/}{mcherculano.github.io} }}
	\date{}
	
	\maketitle
	
	\abstract
	
	\onehalfspacing
	
	 \cite{Frazzini2014} Betting Against Beta (BAB) factor is based on the idea that high beta assets trade at a premium and low beta assets trade at a discount due to investor funding constraints. However, as argued by \cite{Campbell2004} beta comes in "good" and "bad" varieties. While gaining exposure to low-beta, BAB factors fail to recognize that such a portfolio may tilt towards bad-beta. We propose a \textit{Betting Against Bad Beta} factor, built by double-sorting on beta and bad-beta and find that it improves the overall performance of BAB strategies though its success relies on proper transaction cost mitigation. 
	 
	 \vspace{0.5cm}
	 \noindent \textbf{Keywords:} Betting Against Beta, Asset Pricing, Factor Models.
	 
	 \noindent \textbf{JEL Classification:} G12, G14, G11.
	 
	 \newpage
	\section{Introduction}

	In their influential article, \cite{Frazzini2014} propose a \textit{Betting Against Beta} (BAB) factor that exploits the well known \textit{low beta} anomaly related to the empirical failure of the Capital Asset Pricing Model (CAPM). Its basic premise is fairly simple - because low (high) beta stocks earn on average higher (lower) average returns than what the CAPM implies, a BAB factor that is long low-beta assets and short high-beta assets delivers positive CAPM alphas. The authors argue that in theory, the key mechanism driving this result are funding constraints of investors. Instead of investing in the market portfolio and using leverage to tune it to their risk preferences, investors suboptimally overweight risky assets due to leverage constraints. As a consequence, high-beta assets trade at a premium and low-beta assets trade at a discount. 
	
	In the CAPM, beta measures the risk of each stock with respect to the market portfolio. An important outstanding question is whether market betas are driven by cash-flow shocks, associated to firm fundamentals or discount-rate shocks, associated to how aggressively investors discount future cash-flows \citep{Campbell1993}. A growing literature approaches this question using a modified version of \cite{Merton1973} Intertemporal CAPM (ICAPM), in which investors are more concerned with permanent changes in cash-flows rather than temporary movements in discount rates within the aggregate stock market (see \cite{Campbell2004, Campbell2010, Campbell2023}). In the ICAPM, the required return on a stock is determined by two separate betas, one related with permanent cash-ﬂow news to the market, and the
	other with temporary shocks to market discount rates. Hence, \cite{Campbell2004} call betas associated with cash-flow news "bad" betas and those with negative discount-rate news "good" betas, since they correspond to permanent and transitory shocks to investor wealth. 
	
	Against this background, one natural question to ask is whether \textit{Betting Against Beta} can be improved upon by distinguishing between "good" and "bad" beta. We propose a novel \textit{Betting Against Bad Beta} (BABB) factor, built by double-sorting on beta and "bad" beta. Following \cite{Novy-Marx2022} who show that the BAB can be replicated by equal-weight sorting on beta we combine \cite{Frazzini2014} definition of beta with the "bad" beta which is calculated according to \cite{Campbell2004}. The key premise for \textit{Betting Against Bad Beta} is that low-beta assets are often high bad-beta and therefore vulnerable to cash-flow shocks which have permanent effects to net worth. Hence, gaining exposure to low-beta/low-bad beta and shorting high-beta/high-bad beta should in theory improve profitability of the long-short BAB factor. 
	
	The BABB performs well in our backtesting exercise with monthly data and monthly rebalancing on the universe of stocks in Center for Research in Security Prices (CRPS) from 1963:01-2021:01. It yields a moderately higher Sharpe Ratio, which increases by approximately 800 bps as compared to BAB, and significantly higher alphas irrespective of the factor model benchmark according to which they are computed. Nevertheless, because it is constructed in a similar way to the BAB, the observations in \cite{Novy-Marx2022} also apply to the BABB. Both factors are naturally quite similar, however the BABB tilts more strongly towards size and away from profitability and investment. It is thus more expensive to trade, having higher trading costs but similar leverage. With regards to transaction costs, we follow closely \cite{Chen_Velikov_2023, Novy_Marx_Velikov_2024} and consider the simple average of four low-frequency (LF) effective spreads calculated from daily CRPS data and turnover. High-frequency spreads are only available for more recent periods and therefore don't cover our full sample. Nevertheless, we consider LF spread to be good enough proxies in our setting since they have been found to be upward biased \citep{Chen_Velikov_2023}, resulting in more conservative net returns. In addition, recent papers in the anomalies literature have used exclusively LF spreads \citep{Novy_Marx2016, DeMiguel2020}.

	The Betting Against Bad Beta (BABB) factor has a gross mean annualized return of 15.0\% and a annualized volatility of 13.8\% which compares with 11.4\% and 11.3\% of BAB, respectively. The Sharpe ratio of the BABB is 1.09, approximately 800bps higher compared with the BAB which is 1.01 for this specific sample period and construction scheme. After accounting for trading costs, the BABB still rewards investors with higher returns and a greater alpha. The five-factor alpha of the BABB factor is 75 bps per months, in contrasts to 51 bps for our replication of the BAB, similar to the figure presented by \cite{Frazzini2014} of 55 bps which does not however factor in trading costs and is estimated on a different sample. We find that the BABB tilts towards size and away from profitability and investment, as compared to our replications of the BAB. Given its small-cap tilt, it is not surprising that it is more expensive to trade. Performance measures could potentially benefit from appropriate transaction costs mitigation along the lines of \cite{Chen_Velikov_2023}, which we do not consider. Though we find that the Shape Ratios of BAB and BABB factors are sensitive to the way of calculating the beta used in univariate and bivariate sorts, the BABB factor performs competitively as compared to the BAB across alternative beta statistics. 
	
	The remainder of the paper is organized as follows. Section \ref{sec2} outlines the theory and econometric approach underpinning the calculation of \textit{good} and \textit{bad} betas. Section \ref{sec3} makes the case for the novel \textit{Betting Against Bad Beta} factor proposed and explains its construction. Section \ref{sec4} looks into how sensitive the results are to alternative ways of calculating betas and how transaction costs compromise performance. Section \ref{sec5} concludes.

	\section{Good Beta, Bad Beta}
	\label{sec2}
	\subsection{Theory}
	A fundamental question in asset pricing is to understand the relative importance of different types of news in driving stock returns. A large literature combines the log-linearization of \cite{Campbell1988} with a VAR approach as in \cite{Campbell1991} to decompose stock return variance into cash-flow (CF) and discount rate (DR) news. Using the Campbell-Shiller (CS) accounting identity and first-order conditions of a long-term investor with Epstein-Zin preferences, \cite{Campbell1993}, \cite{Campbell2004}, \cite{Campbell2013} and \cite{Campbell2018} derive an approximate closed-form solution for Intertemporal CAPM’s risk prices. The one-period innovation to the Stochastic Discount Factor (SDF) for the ICAPM can therefore be written as a function of news. Start by writting the one-period innovation of the log SDF as follows:
	
	\begin{equation}
		m_{t+1}-E_t m_{t+1} = \dfrac{\theta}{\psi} \big[h_{t+1} - E_t h_{t+1}\big] + \gamma \big[r_{t+1} - E_t r_{t+1}\big], \label{SDF_1}
	\end{equation}
	
	where $h_{t+1}=ln(W_{t+1}/C_{t+1})$ denotes the future value of a consumption claim, measured as the ratio of future wealth $W_{t+1}$ by future consumption $C_{t+1}$ and captures the effects of intertemporal hedging on asset prices. $\gamma$ represent investor risk aversion and $\theta$ is a behavioral parameter which is a function of $\gamma$ and of the Elasticity of Intertemporal substitution $\psi$. The log return on wealth is constructed as $r_{t+1}=ln(W_{t+1}/(W_t-C_t))$, measuring the log value of wealth tomorrow divided by reinvested wealth today.
	
	The SDF (\ref{SDF_1}) can be expressed as a function of news by imposing asset pricing restrictions. We obtain:
	
	\begin{align}
		m_{t+1}-E_t m_{t+1} = -\gamma [r_{t+1}-E_tr_{t+1} ]-(\gamma -1)N_{DR,t+1} = \\
		= -\gamma N_{CF,t+1}+N_{DR,t+1}, \label{SDF_2}
	\end{align}
	
	yielding the log SDF in terms of the market return and news about future variables. $N_{CF}$, $N_{DR}$ follows \cite{Campbell2004} notation for cash-flow and discount rate news which can be interpreted as permanent and transitory shocks to investors' wealth. Cash-flow news-driven returns are never reverted, while returns driven by discount rate news mean-revert. \cite{Campbell2018} allows for a third priced risk factor driving the SDF, capturing revisions in expectations of future risk, denoted by $N_{RISK}$. This term represents news about the conditional variance of returns plus the SDF, $Var_t(r_{t+1}+m_{t+1})$. However, it cancels out in an homoskedastic ICAPM world and we leave it out of the analysis. Hence we can write:
	\begin{align}
		N_{CF,t+1}=r_{t+1}-E_t r_{t+1} + N_{DR,t+1}.
	\end{align}

	\cite{Campbell2004} build on such a framework and define cash-flow and discount-rate betas as
	
	\begin{equation}
		\beta_{i,CF}= \dfrac{Cov(r_{i,t},N_{CF,t})}{Var(r_{M,t}-E_{t-1}r_{M,t})}, \label{beta_bad}
	\end{equation}
	and
	\begin{equation}
		\beta_{i,DR}= \dfrac{Cov(r_{i,t},N_{DR,t})}{Var(r_{M,t}-E_{t-1}r_{M,t})}, \label{beta_good}
	\end{equation}
	with 
	
	\begin{equation}
		\beta_i = \beta_{CF}+\beta_{DR}, \label{beta} 
	\end{equation}
	
	for given asset $i$. The authors call CF betas "bad betas" because investors demand a high price to bear this risk, and DR betas "good betas". 
		
	\subsection{Econometric Approach}
	
	Assume that the economy is well described by a first-order vector-autoregressive (VAR) model of the form
	
	\begin{equation}
		x_{t+1} = \mu^x+\Gamma x_t +\sigma u_{t+1}. \label{VAR} \tag{6}
	\end{equation}
	
	Where $x_t$ is a $n \times 1$ vector of state variables that include $r_t$ (excess stock market return) as its first element and $n -1$ other variables that help to measure or forecast excess returns. $\mu^x$ and $\Gamma$ are an $n \times 1$ vector and an $n \times n$ matrix of constant parameters, and $u_{t+1}$ is a vector of shocks to the state variables. In general, an homoskedastic VAR is assumed and therefore $\sigma$ in ($\ref{VAR}$) is time-invariant. Given this VAR structure and assuming homoskedasticity, news about cash-flow and discount rates can be retrieved from the model as follows
	
	\begin{equation}
		N_{DR,t+1} = (E_{t+1}-E_{t}) \sum_{j=1}^{\infty} \rho  r_{t+1+j} \\
		= e_1' \rho \Gamma (I- \rho \Gamma)^{-1} \sigma u_{t+1},
	\end{equation}
	
	while the implied cash-flow news is obtained as 
	
	\begin{equation}
		N_{CF,t+1} = (r_{t+1}-E_{t}r_{t+1}) + N_{DR,t+1}\\
		= \big(e_1'+ e_1'\rho \Gamma (I- \rho \Gamma)^{-1} \big) \sigma u_{t+1},
	\end{equation}
	
	where $\rho$ is a log-linearization parameter in the Campbell-Shiller identity, set to 0.95 by most authors. We follow \cite{Campbell2004} and include four state variables in the model: excess stock market returns $r_t$, (measured as the log
	excess return on the CRSP value-weighted index
	over Treasury bills); the yield spread, between 10 year and 3 months Treasury Bonds, the Cyclically Adjusted Price Earning Ratio (CAPE) and the small-stock	value spread (measured as the difference between the log book-to-market ratios of small
	value and small growth stocks). 
	
	The motivation for choosing this set of predictions are three-fold. First, the yield curve is an indicator of the business cycle, and it's plausible that stock market returns fluctuate with the business cycle for various reasons. Second, if price-earnings ratios are high, it logically leads to lower long-term expected returns, assuming earnings growth remains unchanged. Thirdly, the difference in value between small growth and small value stocks is explained by the ICAPM. Specifically, if small growth stocks are expected to yield lower returns and small value stocks higher returns, and these differences are not accounted for by CAPM betas, then ICAPM suggests that returns from small growth stocks will predict lower, and those from small value stocks will predict higher future market returns. This set of variables is also used by \cite{Campbell2004}. 
	
	Additional theories also link the small-stock value spread to overall market discount rates. For example, small growth stocks, which are expected to generate cash flows further in the future, may have prices that are more affected by changes in discount rates, similar to how long-duration bonds are more sensitive to interest rate changes. Moreover, small growth firms may rely more on external financing, making their valuations more susceptible to the conditions of the equity market and broader financial environment. Lastly, periods of irrational investor optimism might particularly impact the valuation of small growth stocks.
		
\begin{figure}[H]
	\centering
	
	\caption{Time-Series of Discount Rate and Cash-Flow News}
	\includegraphics[width=0.8\textwidth]{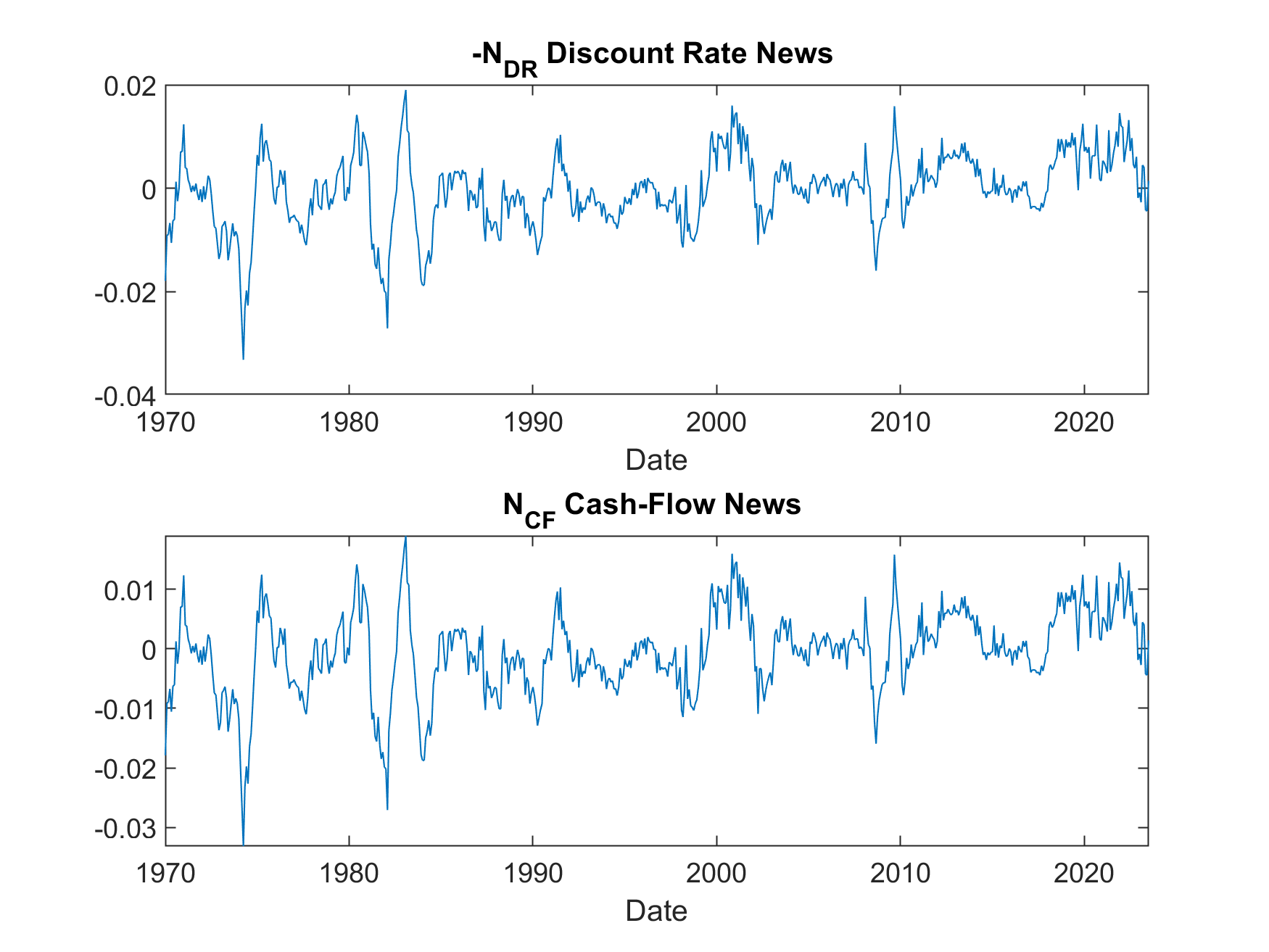} 
	\label{fig:image1} 
	
	\begin{tablenotes}
		\small
		\item \textbf{Notes:} News shocks are calculated as outlined in equations (8)-(9), following \cite{Campbell2004} for the full CRSP sample until 2023:07.
	\end{tablenotes}
\end{figure}

Figure \ref{fig:image1}	show the time-series of discount rate and cash-flow news, which allow us to calculate "good" and "bad" betas, as defined in equations \ref{beta_good} and \ref{beta_bad}, the starting point to our exercise. Next, we outline how these two betas can be used to construct the novel \textit{Betting Against Bad Beta} factor. 
		
	\section{Betting Against (Bad) Beta}
\label{sec3}
Since its publication, the \textit{Betting Against Beta} (BAB) factor of \cite{Frazzini2014} has become one of the most widely known Asset Pricing anomalies in the literature, associated to defensive and low volatility investment strategies. The authors argue that in theory, the key mechanism driving this result are funding constraints of investors. Instead of investing in the market portfolio and using leverage to tune it to their risk preferences, investors suboptimally overweight risky assets due to leverage constraints. As a consequence, high-beta assets trade at a premium and low-beta assets trade at a discount.

To exploit this anomaly, the \cite{Frazzini2014} BAB factor is long a portfolio of low-beta assets and short a portfolio of high-beta assets. The long-short factor is then hedged to achieve market neutrality by leveraging low-beta and deleveraging high-beta using these portfolios' predicted betas with a view of netting out the scaled portfolio betas which is then equal to zero resulting in the following identity for BAB returns

\begin{equation}
	r_{BAB} = \beta_L^{-1} (r_L - r_f) - \beta_H^{-1} (r_H - r_f),
\end{equation} 
where $r_L$, $r_H$ and $r_f$ are the returns of the low-beta, high-beta portfolio, and "risk-free" short-term treasuries, respectively, and $\beta_L$, $\beta_H$ are the estimated betas of the low and high beta portfolios, plotted below.

As noted by \cite{Novy-Marx2022}, though in the original paper the BAB factor is build using non-standard procedures, it can be replicated via an equal-weighted portfolio based on a quantile-sort on beta by holding the top and bottom thirds of stocks selected on the basis of the Frazzini-Pederson betas \footnote{see Figure 2 on pp.84 in their paper.} which are calculated for a given asset $i$ by the authors as follows

\begin{equation}
	\beta^i_{FP} = \Bigg(\dfrac{\rho^i_{5}}{\rho^i_{1}}\Bigg) \beta^i_{1}, \label{betafp}
\end{equation}

where $\rho^i$ denotes the asset’s correlation with the market, $\beta^i$ is its beta estimated from a CAPM regression, with subscripts denoting estimation windows measured in years. Here betas are estimated using one-year rolling standard deviation for volatilities and a five-year horizon for the correlation to account for the fact that correlations move more slowly than volatilities. 

\begin{figure}[H]
	\centering
	
	\caption{Average betas of top and bottom terciles used to construct the BAB factor. }
	\includegraphics[width=0.4\textwidth]{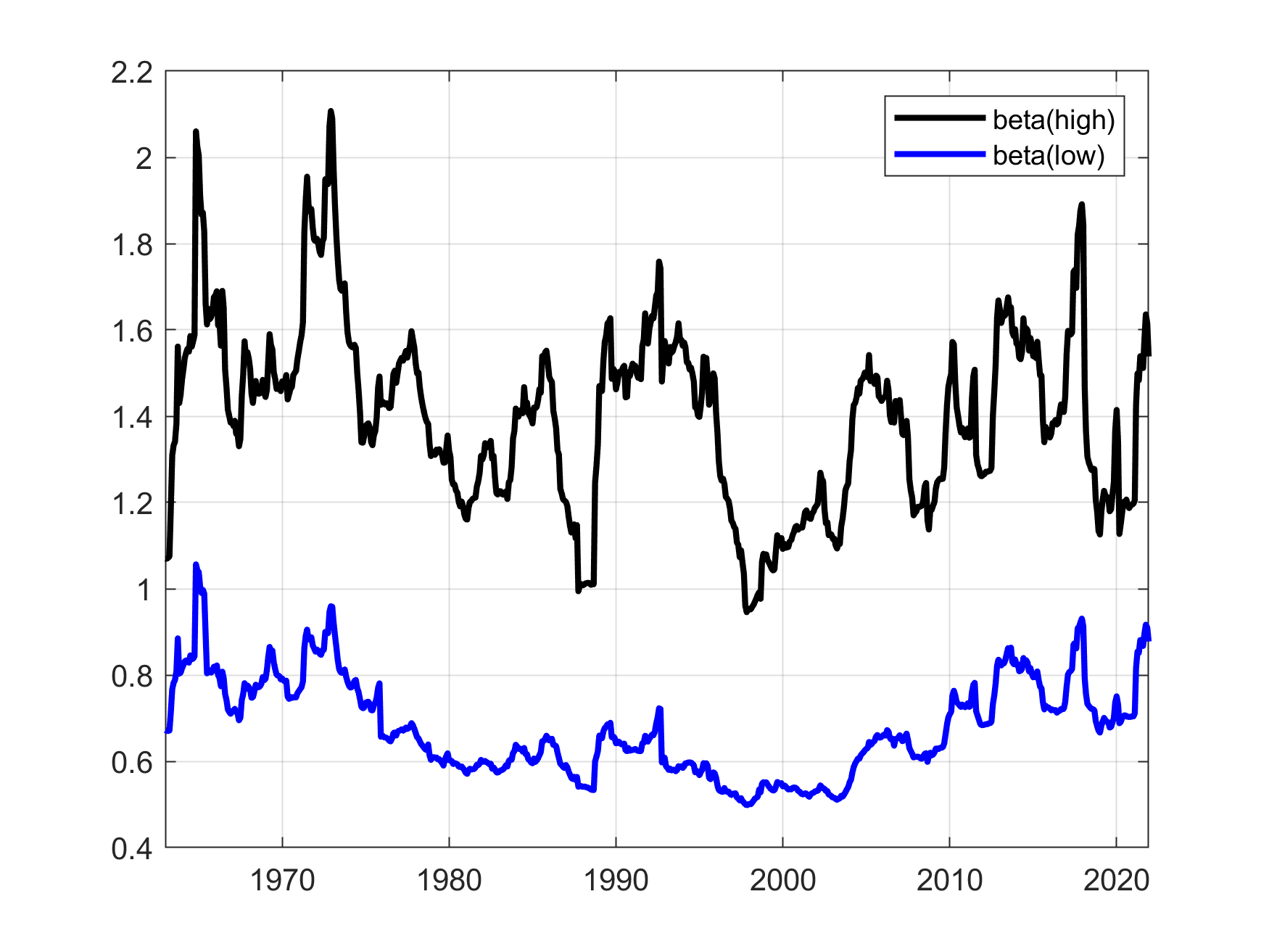} 
	\includegraphics[width=0.4\textwidth]{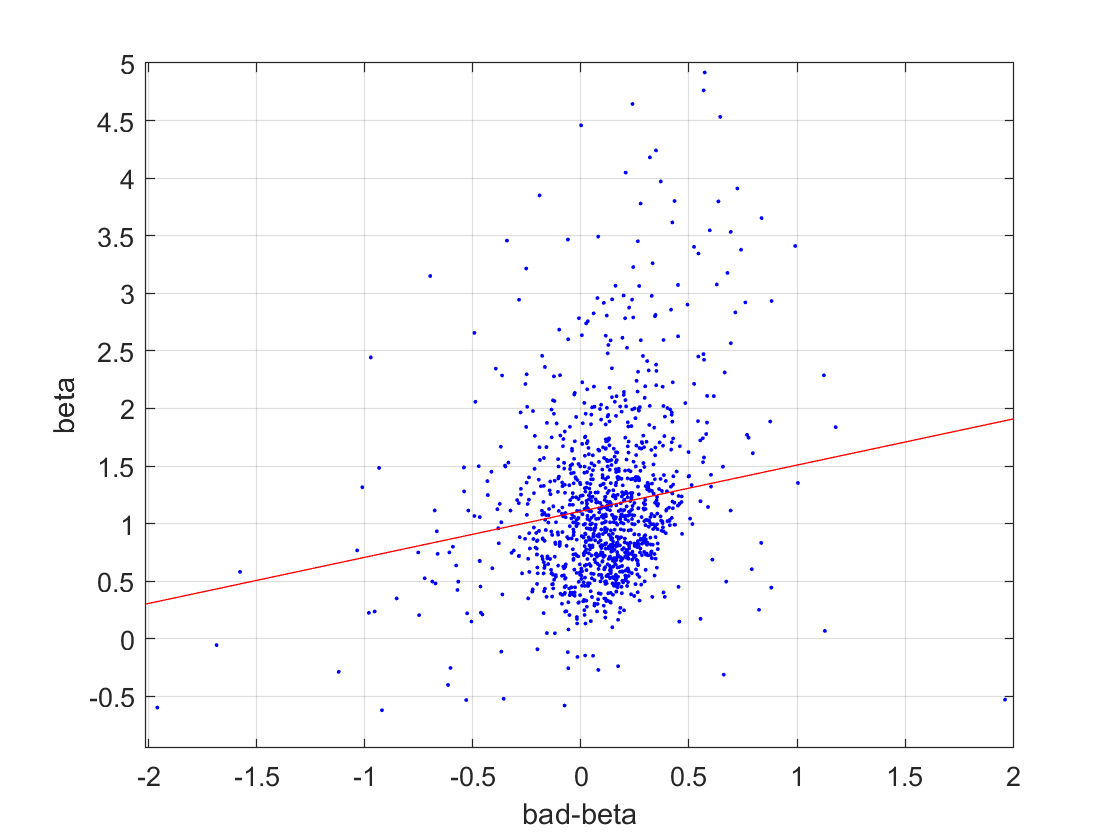} 
	\label{fig:image2} 
	
\end{figure}

Figure 2 shows the wedge between the average beta of the high and low terciles used to construct the BAB. It also shows for each firm in the CRPS database in use, the average beta and bad-beta over the sample considered. The key message here is that, even though beta and bad-beta are linearly related, the link is not strong. Thus, low-beta stocks often have high bad-betas.

The main premise for building a \textit{Betting Against Bad Beta} factor is that, assets with high "bad" betas are more vulnerable to cash-flow news shocks that, in theory, have permanent effects as compared to discount rate news shocks which are thought to be transitory, therefore capturing mean-reverting dynamics in stock prices. Therefore, a BABB strategy tilting the low-beta portfolio away from "bad" beta and shorting high beta which is also high "bad" beta, ought to deliver better risk-adjusted return.  

\subsection{BABB factor construction and performance}

To understand the prospect of such a strategy, we build an equal-weighted $3 \times 3$ double-sort portfolio on beta $\beta$ and bad beta $\beta_{CF}$ and compare it with a univariate sort on $\beta$ which we call BAB throughout the paper \footnote{Even though \cite{Frazzini2014} construct the BAB via rank-ordering, \cite{Novy-Marx2022} show that an equal-weighted tercile univariate sorting is a good proxy. }. The BABB returns are then given by

\begin{equation}
	r_{BABB} = \beta_{LL}^{-1} (r_{LL} - r_f) - \beta_{HH}^{-1} (r_{HH} - r_f),
\end{equation}

where $r_{LL}$, $r_{HH}$ and $r_f$ are the returns of the low-beta/low-bad-beta, high-beta/high-bad-beta portfolios, and "risk-free" short-term treasuries, respectively, and $\beta_{LL}$, $\beta_{HH}$ are the estimated betas of the low/low and high/high beta/bad-beta portfolios. Both factors are built on the entire sample of US stocks in the CRPS dataset for the sample 1963:01-2021:12. We work with monthly data and monthly rebalancing for backtesting purposes. $\beta$ and $\beta_{CF}$ are calculated in real-time, following equations (\ref{betafp}) for $\beta$ and (\ref{beta_bad})-(\ref{beta_good}) are used to calculate cash-flow and discount rate news on an expanding window of data. Cash-flow news is then used to implement real-time bad-beta following equation (\ref{beta_bad}) where a 3-year rolling-window of stock returns are used. 

In the Figures below, we compare the performance of our BAB replication with our novel \textit{Betting Against Bad Beta} (BABB) factor. Trading costs are considered following \cite{Chen_Velikov_2023, Novy_Marx_Velikov_2024} by using an effective firm-level bid-ask spread proxy formed by the equal-weighted average of four low-frequency (LF) used by the authors. \footnote{We provide more details about how we compute trading costs and will look more carefully at their implication for the BABB in section \ref{Robustness}.}

\begin{figure}[H]

	\centering
	\caption{Cumulative returns of the Betting Against (Bad) Beta factor.}
	\adjustbox{center=\textwidth}{\includegraphics[width=1.2\textwidth]{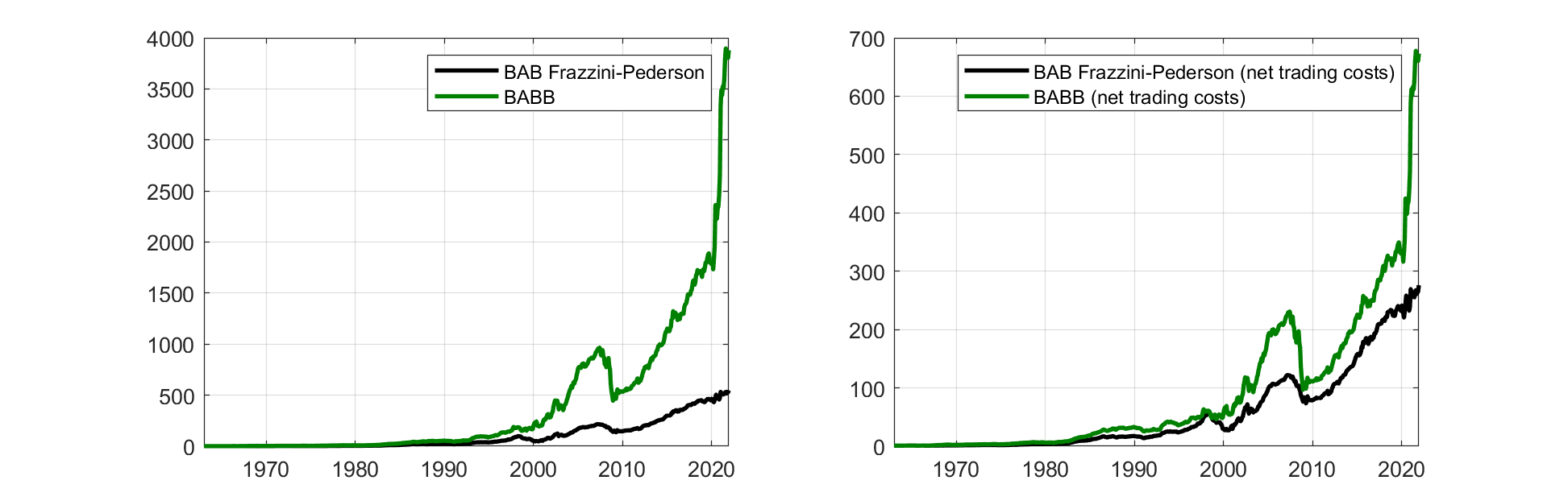}} 
	\label{fig1} 
	
	\begin{tablenotes}
		\item \textbf{Notes:} \small  The BAB is long-short the bottom-top thirds of equal-weighted US stocks in CRPS, sorted on the basis of the Frazzini-Pederson betas from 1963:01-2021:12. The BABB factor is built following the same procedure but double-sorting on betas and also on "bad" betas constructed according to \cite{Campbell2004}. Rebalancing is monthly, RHS panel assumes trading costs, computed using an average Low-Frequency effective firm-level bid-ask spread as in \cite{Chen_Velikov_2023}. 
	\end{tablenotes}
\end{figure}

\begin{figure}[H]
	\centering
	 
	\caption{Risk-Return profile of Betting Against Beta Factors.}
	\includegraphics[width=0.5\textwidth]{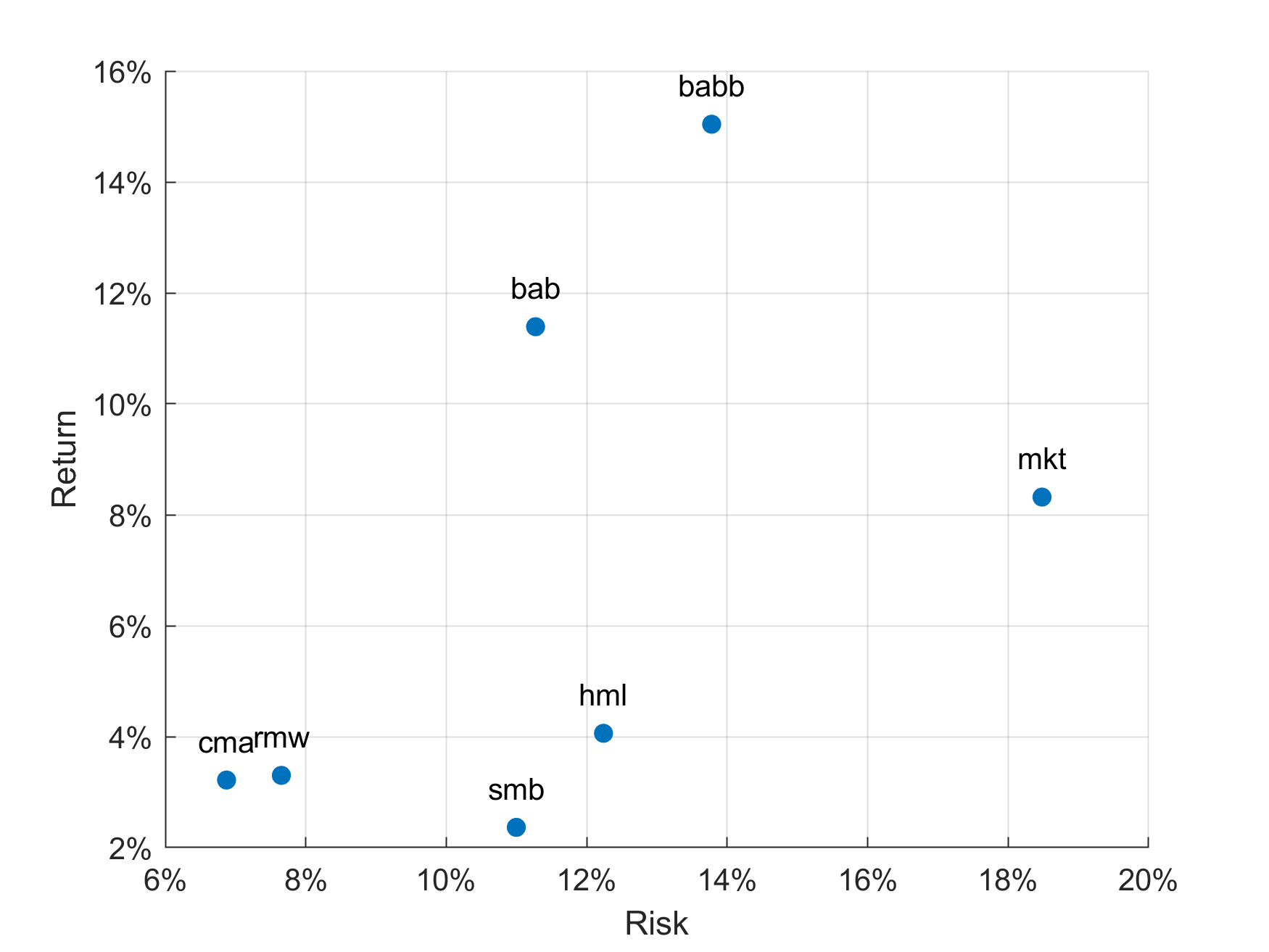} 
	\includegraphics[width=0.5\textwidth]{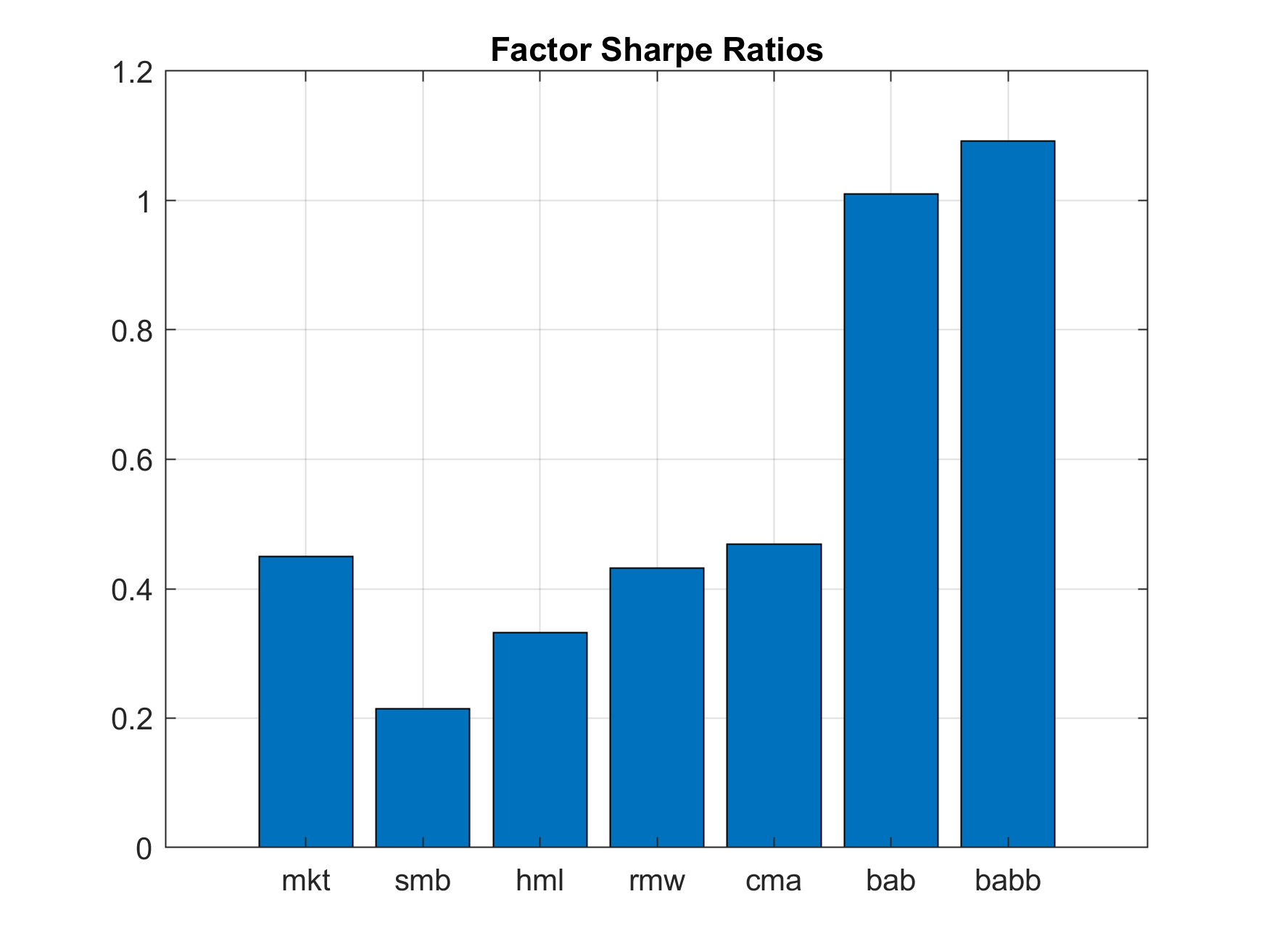}
	\begin{tablenotes}
		\item \textbf{Notes:} The 5 Fama-French factors are: mkt (market risk factor), smb (Small minus Big), hml (High minus Low), rmw (Robust Minus Weak) and the cma (Conservative Minus Aggressive) to which we add BAB (Betting Against Beta) and BABB (Betting Against Bad Beta). Risk and return are calculated at a monthly frequency and annualized with a sample from 1963:01-2021:12.
	\end{tablenotes}\label{fig4}
\end{figure}

The Betting Against Bad Beta (BABB) factor has a gross mean annualized return of 15.0\% and a annualized volatility of 13.8\% which compares with 11.4\% and 11.3\% of BAB, respectively. The Sharpe ratio of the BABB is 1.09, approximately 800bps higher compared with the BAB which is 1.01 for this specific sample period and construction scheme. Figure \ref{fig4} puts these statistics into perspective by comparing them with the 5 factors of \cite{FAMA20151}. We find that, in our sample, Sharpe Ratios of both the BAB and BABB factors are at least twice those of the 5 FF factors. Albeit, both the BAB and BABB are significantly riskier. For instance, the BABB's annualized volatility of returns is twice that of the market portfolio.

To understand the underlying drivers of risk and return of the BABB factor and how it compares to the BAB, we regress these factor returns on the 5 Fama-French factors plus momentum. In Table \ref{t1}, Panels A and C show the loadings of each factor on the returns of BAB and BABB, respectively. A number of key differences are salient. First, exposure to these factors explain at most 11 per cent of BABB returns, which contrasts with roughly 25 per cent in the case of BAB. Both BAB and BABB have statistically significant alphas but BABB alphas are substantially higher. In particular, we find that the BABB has a five-factor alpha of 75 bps per month, which compares to 51 bps of our replication of \cite{Frazzini2014}. These results are comparable with the result of 55bps per month reported in the original paper, albeit for a different sample, running from January 1963 until March 2013.

In column (5), which reports the 6 factor regression results, one can see that the alpha coefficient for the BABB is twice that of BAB, a result which seems to hold regardless of whether trading costs are considered. It is also worth noting that, while coefficients of all factors are statistically significant for BAB returns, the market factor doesn't seem to be relevant in explaining BABB returns. Some differences in the magnitude of the coefficients are also interesting. For instance, the coefficient on the size factor is more than twice greater for BABB returns, suggesting that BABB aggressively tilts towards smaller firm equity. This is in contract to the BAB that tilts towards profitability and investment. These finding helps rationalize the fact that, when accounting for trading costs, the wedge between cumulative BABB and BAB results decreases significantly. Smaller firms tend to be more expensive to trade, having greater bid-ask spreads. 	Tables \ref{t2} and \ref{t3} provide a more granular view of these results, with factor regressions of each portfolio sort.

\newpage

\section{Robustness} \label{Robustness}
\label{sec4}
\subsection{Different ways of calculating beta}

Betting Against Beta strategies rely on the specific methodology of calculating individual stocks’ betas as well as the stability those estimates.  The \cite{Frazzini2014} approach calculates beta using a combination of correlations and volatilities, where the correlation between a stock's returns and the market's returns is estimated over a five-year horizon, and the volatility is estimated using one year of daily data. This method is intended to capture the systematic risk of stocks while being responsive to short-term market conditions. However, \cite{Novy-Marx2022} argue that this method introduces significant biases into the beta estimates, particularly because it combines data from two different time horizons. The longer five-year period used for correlation estimation tends to smooth out short-term fluctuations, potentially underestimating the true beta during volatile periods. Conversely, the one-year period used for volatility estimation is more sensitive to recent market conditions, which can lead to an overestimation of beta in times of market turbulence. This mismatch can create a distorted picture of a stock's systematic risk and may lead to unstable beta estimates that are heavily influenced by the chosen time period. In particular, during periods of high volatility, the beta estimates can become excessively noisy, reducing the effectiveness of the BAB strategy in maintaining market neutrality. 

In response to these concerns, we follow \cite{Novy-Marx2022} and estimate individual stock betas using six different procedures and reconstruct the BABB strategy using these alternative estimates. These include betas calculated via simple OLS regression with one year of either daily returns or three-day overlapping returns, \cite{DIMSON1979197} corrected estimates employing one lag and the level of shrinkage used by \cite{Frazzini2014}, that accounts for non-synchronous trading, the beta by \cite{Welch2019} which is similar to the OLS beta but uses winsorized returns, \cite{Vasicek1973} betas which applies shrinkage and what \cite{Novy-Marx2022} consider standard, obtained by shrinking the Dimson corrected beta towards 1. 

\begin{figure}[H]
	\centering
	\caption{BABB Sharpe Ratio with different betas}
	\includegraphics[width=0.75\textwidth]{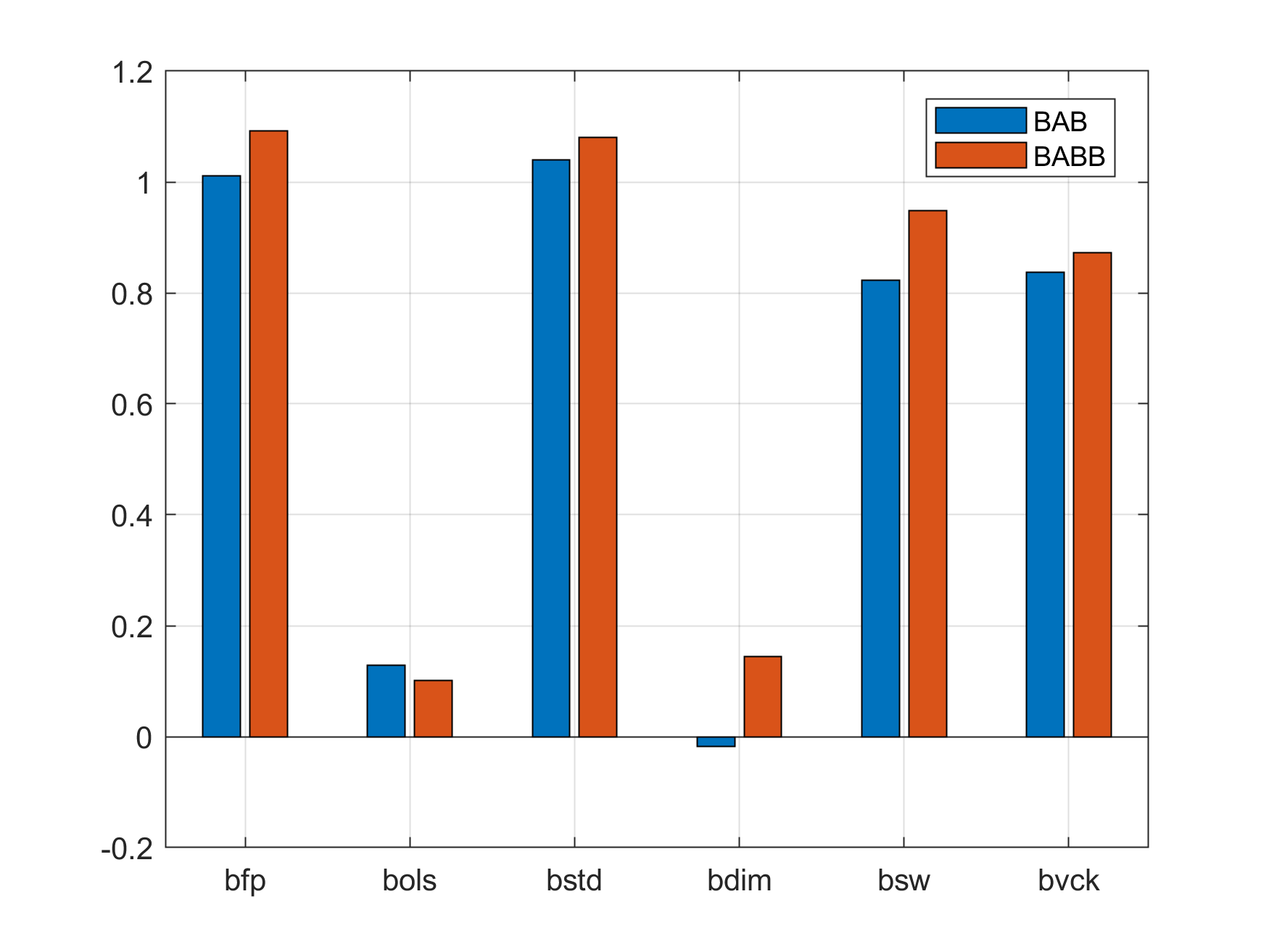} 
	
	\begin{tablenotes}
		\small
		\item \textbf{Notes:} Sharpe Ratios are calculated on a monthly basis and annualized. The BAB and BABB strategies are reconstructed with six alternative beta signals: ``bfp`` \cite{Frazzini2014}, ``bols`` obtained by simple OLS regression, ``bdim`` \cite{DIMSON1979197}, ``bvck`` \cite{Vasicek1973} and ``bst`` Dimson corrected betas shrunk to 1.  
	\end{tablenotes}
	\label{fig5}
\end{figure}

Figure \ref{fig5} shows Sharpe Ratios of the BAB and BABB strategies based on different beta estimates. While it is true that the performance is sensitive to the beta used, we find that in five out of the six alternative beta estimates Sharpe Ratios where higher for BABB as compared to BAB.

\subsection{Leverage and Transaction Costs}

Similar to the BAB of \cite{Frazzini2014}, the BABB is constructed by leveraging the low beta portfolio and deleveraging the high beta portfolio, with a view of achieving market neutrality. \cite{Novy-Marx2022} argue that this procedure amplifies exposure to small-cap stocks, increases transaction costs, and introduces significant liquidity risks. With regards to transaction costs, the authors also emphasize how these costs can significantly erode the strategy's theoretical returns and raise concerns about its practical implementation. To understand how BABB leverage compares with that of BAB, we compute $L=B_{H}/B_{L}$ and plot the time-series in Figure \ref{fig6} below. We observe that the BABB factor leverage is not remarkable distinct from that of the BAB. 

\begin{figure}[H]
	
	\caption{Leverage of BAB and BABB factors.}
	\centering
	\includegraphics[scale=1]{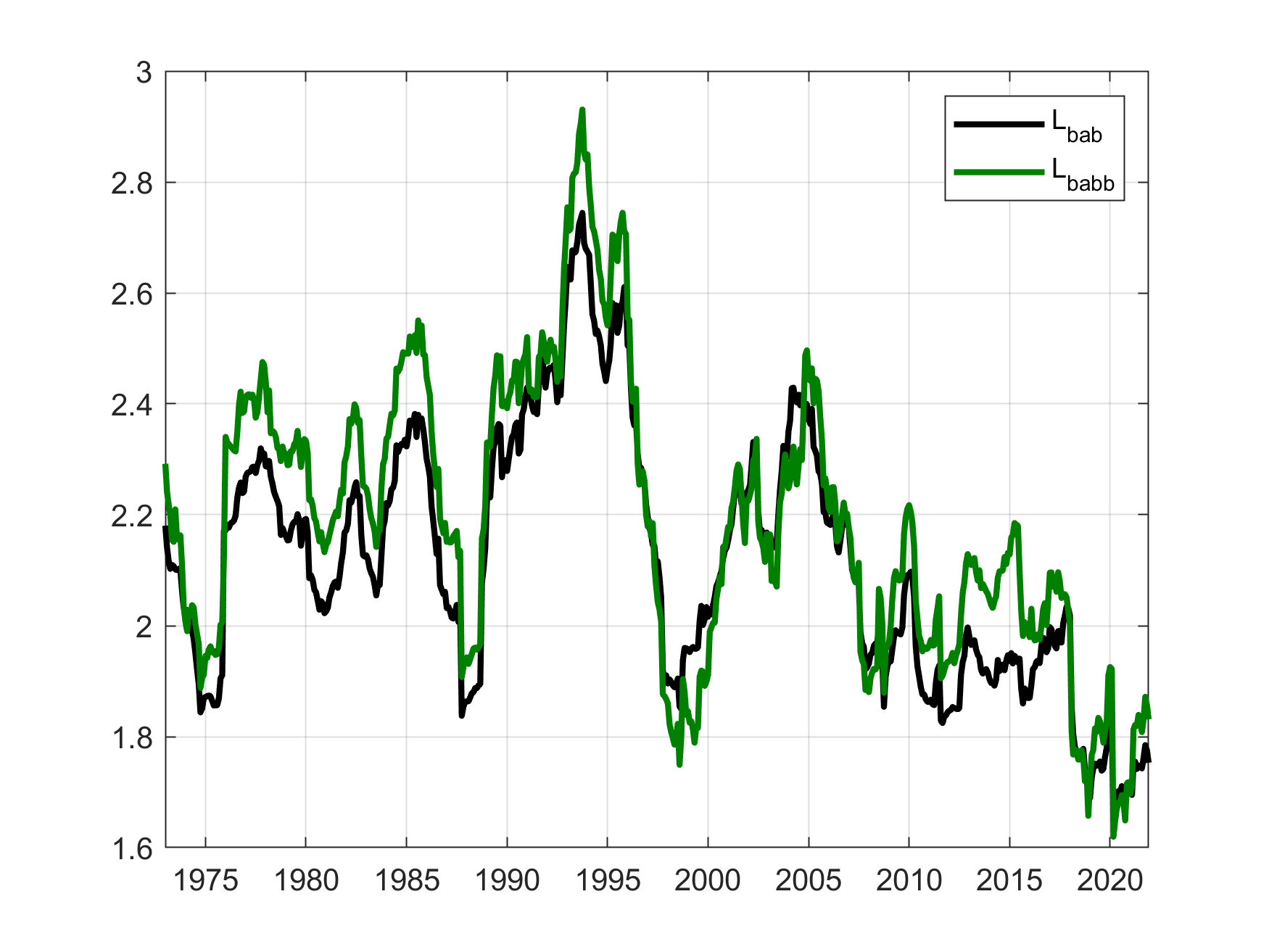} 
	\begin{tablenotes}
		\small
		\item \textbf{Notes:} Leverage is defined by taking the ratio $\beta_{H}/\beta_{L}$ in case of the BAB or $\beta_{HH}/\beta_{LL}$ for the BABB.
	\end{tablenotes}
	\label{fig6}
\end{figure}

With regards to transaction costs, throughout the paper we work with BAB and BABB factor returns gross and net of transaction costs. Next, we explain how these are calculated. We take after \cite{Chen_Velikov_2023,Novy_Marx_Velikov_2024} and consider the simple average of four low-frequency (LF) measures of transaction costs, based on half the effective firm-level bid-ask spread. Specifically, we consider \cite{HASBROUCK2009} Gibbs estimate, \cite{CORWIN2012} high-low spread, \cite{Ranaldo2017} close-high-low spread , and  \cite{Kyle2016} invariance-based volume-over-volatility measure (VoV). LF transaction cost measures are computed with daily CRSP data and proxy for their high-frequency counterparts, which use intraday data. Even though LF measures are found to be upward biased by about 25–50 bps post-2005 \citep{Chen_Velikov_2023}, they are available for our sample and have been extensively used in the literature on market anomalies. 

\begin{table}[H]
	\centering
	\caption{Average Firm Size, Number of Stocks, Trading Costs, and Turnover for BAB and BABB Portfolios}
	\medskip 
	\begin{adjustbox}{scale=0.9, center}
		\begin{tabular}{|l|l|l|l|l|}
			\hline
			Portfolio & Avg Firm Size ($10^6\$$) & Avg No. Stocks & tcosts & turnover \\ \hline
			\textbf{BAB} & ~ & ~ & ~ & ~ \\ \hline
			1 (low beta) & 2082 & 1156 & 0.0017 & 0.1642 \\ \hline
			2 & 2962 & 1162 & 0.0023 & 0.2525 \\ \hline
			3 (high beta) & 2086 & 1162 & 0.0023 & 0.2154 \\ \hline
			\textbf{BABB} & ~ & ~ & ~ & ~ \\ \hline
			1 (low beta/low bad beta) & 1178 & 457 & 0.0031 & 0.3261 \\ \hline
			2 & 2787 & 380 & 0.0034 & 0.4248 \\ \hline
			3 & 1779 & 268 & 0.0041 & 0.3995 \\ \hline
			4 & 1976 & 340 & 0.0040 & 0.4353 \\ \hline
			5 & 3908 & 427 & 0.0033 & 0.4805 \\ \hline
			6 & 2890 & 363 & 0.0034 & 0.4174 \\ \hline
			7 & 1396 & 325 & 0.0050 & 0.4229 \\ \hline
			8 & 3289 & 315 & 0.0049 & 0.5505 \\ \hline
			9 (high beta/high bad beta) & 2115 & 491 & 0.0034 & 0.3567 \\ \hline
		\end{tabular}
	\end{adjustbox}
	\begin{tablenotes}
	\small
	\item \textbf{Notes:} The average firm size refers to each porfolio's sort market capitalization divided by the number of firms in that portfolio. tcosts refer to trading costs, computed as the simple average of four low-frequency (LF) measures of transaction costs (i.e. Gibbs estimate, high-low spread, close-high-low spread  and the  invariance-based volume-over-volatility measure), based on half the effective firm-level bid-ask spread. Turnover is the average change in weights of stocks held in the corresponding portfolio, throughout the sample period. 
	\end{tablenotes}
	\label{t3}
\end{table}

We find that BABB factor trading costs are significantly higher compared to those of BAB. Transaction costs of the long-leg of the BABB factor porfolio are nearly twice as expensive, while they are almost 50\% more expensive for the short-leg. Turnover is also consistently higher over each of the $3 \times 3$ double sort, compared to the univariate sort used to construct the BAB. This is consistent with the finding previously discussed that the BABB seems to tilt towards size and away from profitability and investment. Nevertheless, we observe that even after factoring in transaction costs, the BABB still outperforms historically the BAB returns and delivers higher alpha. This can be observed in the right-hand-side panel of Figure \ref{fig1} and in table \ref{t2}. Though transaction costs close the wedge between cumulative returns of the BAB and BABB, the later still outperforms. A similar result is apparent in table \ref{t2}, where we observe that the five-factor alpha of the BABB decreases by 25 bps to 75 bps, which contrasts with a decrease of 10 bps in case of the BAB. Nevertheless, net of transaction costs, the BABB's alpha is still higher. Transaction costs do however decrease the BABB's performance significantly compared to the BAB on a risk-adjusted basis, emphasizing the importance of a suitable cost mitigation strategies as highlighted by \cite{Chen_Velikov_2023}. 

\section{Conclusion}
\label{sec5}

A novel Betting Agaist Bad Beta (BABB) factor is proposed. The BABB is similar in nature to \cite{Frazzini2014} BAB factor, but also distinguishes between \textit{good} and \textit{bad} beta, shorting the later. This is achieved via a $3 \times 3$ bivariate sort of beta and \textit{bad} beta. We find that the BABB factor improves upon the BAB, delivering higher Sharpe Ratios and greater alpha. The economic argument for why it works is based on the idea that beta comes in good and bad varieties \citep{Campbell2004}. High \textit{bad} beta is related to permanent rather than transitory effects to valuations. Therefore the BABB shorts high beta stocks which have high \textit{bad} beta, while gaining exposure to low beta, low \textit{bad} beta equity that is less vulnerable to persistent valuation shocks. 

		\newpage
		\setlength\bibsep{10pt}
		\bibliography{library}        
		\bibliographystyle{apalike}
		
		\appendix

		\newpage 
	
\begin{table}[H] 
	\centering
	\caption{Factor Regressions for the Betting Against (Bad) Beta factors}
	\paragraph{}
	\begin{adjustbox}{scale=0.65, center}
		\begin{tabular}{ l|l l l l l|}
			\multicolumn{6}{l}{\textbf{Panel A: Betting Against Beta}} \\
			
			Explanatory factor & (1) & (2) & (3) & (4) & (5) \\ \hline
			alpha & 0.993*** & 0.876*** & 0.681*** & 0.613*** & 0.47*** \\ \hline
			mkt   & -0.073*** & -0.032 & 0.013 & 0.038 & 0.07** \\ \hline
			smb   & ~         & 0.012 & 0.015 & 0.137*** & 0.13*** \\ \hline
			hml   & ~ & 0.317*** & 0.403*** & 0.101* & 0.204*** \\ \hline
			rmw   & ~ & ~ &  & 0.49*** & 0.454*** \\ \hline
			cma   & ~ & ~ & ~ & 0.458*** & 0.397*** \\ \hline
			umd   & ~ & ~ & 0.227*** & ~ & 0.199*** \\ \hline
			R-squared &  0.01  &  0.09  &  0.16  &  0.20  &  0.25  \\ \hline
			\multicolumn{6}{c}{} \\
			
			\multicolumn{6}{l}{\textbf{Panel B: Betting Against Beta (net transaction costs)}} \\
			& ~ & ~ & ~ & ~ & ~ \\ \hline
			Explanatory factor  & (1) & (2) & (3) & (4) & (5) \\ \hline
			alpha & 0.894*** & 0.777*** & 0.584*** & 0.515*** & 0.372*** \\ \hline
			mkt & -0.073*** & -0.033 & 0.013 & 0.037 & 0.069** \\ \hline
			smb & ~ & 0.013 & 0.016 & 0.138*** & 0.131*** \\ \hline
			hml & ~ & 0.318*** & 0.403*** & 0.102* & 0.204*** \\ \hline
			rmw & ~ & ~ &  & 0.488*** & 0.452*** \\ \hline
			cma & ~ & ~ & ~ & 0.458*** & 0.397*** \\ \hline
			umd & ~ & ~ & 0.225*** & ~ & 0.197*** \\ \hline
			R-squared &  0.01  &  0.09  &  0.16  &  0.20  &  0.25  \\ \hline
			\multicolumn{6}{c}{} \\
			
			\multicolumn{6}{l}{\textbf{Panel C: Betting Against Bad Beta}} \\
			Explanatory factor & (1) & (2) & (3) & (4) & (5) \\ \hline
			alpha & 1.243*** & 1.115*** & 1.003*** & 1.004*** & 0.919*** \\ \hline
			mkt & 0.001 & -0.029 & -0.005 & 0.007 & 0.024 \\ \hline
			smb & ~ & 0.316*** & 0.318*** & 0.353*** & 0.35*** \\ \hline
			hml & ~ & 0.244*** & 0.29*** & 0.084 & 0.141** \\ \hline
			rmw & ~ & ~ &  & 0.208*** & 0.188*** \\ \hline
			cma & ~ & ~ & ~ & 0.251** & 0.217** \\ \hline
			umd & ~ & ~ & 0.123*** & ~ & 0.11*** \\ \hline
			R-squared &  -    &  0.09  &  0.11  &  0.10  &  0.10  \\ \hline
			\multicolumn{6}{c}{} \\
			
			\multicolumn{6}{l}{\textbf{Panel D: Betting Against Bad Beta (net transaction costs)}} \\
			Explanatory factor & (1) & (2) & (3) & (4) & (5) \\ \hline
			alpha & 0.994*** & 0.865*** & 0.756*** & 0.757*** & 0.674*** \\ \hline
			mkt & 0.015 & -0.01 & 0.016 & 0.027 & 0.045 \\ \hline
			smb & ~ & 0.321*** & 0.322*** & 0.358*** & 0.354*** \\ \hline
			hml & ~ & 0.289*** & 0.337*** & 0.125* & 0.184*** \\ \hline
			rmw & ~ & ~ & & 0.202*** & 0.182** \\ \hline
			cma & ~ & ~ & ~ & 0.26** & 0.224** \\ \hline
			umd & ~ & ~ &  0.128*** & ~ & 0.115*** \\ \hline
			R-squared &  -    &  0.09  &  0.10  &  0.10  &  0.11 \\ \hline
		\end{tabular}
	\end{adjustbox}
	\begin{tablenotes}
		\small
		\item \textbf{Notes:} The 5 Fama-French factors are: mkt (market risk factor), smb (Small minus Big), hml (High minus Low), rmw (Robust Minus Weak) and the cma (Conservative Minus Aggressive); to which we add umd (Momentum). Regressions (1) to (5) correspond to the CAPM, 3 Factor Fama-French model, the Carhart Four-Factor Model, the 5 Factor Fama-French Model and the Fama-French Six-Factor Model. The dependent variables are the BAB and BABB factors gross returns and net returns, after accounting for transaction costs. All variables at monthly frequency with a sample from 1963:01-2021:12.
	\end{tablenotes}
	\label{t1}
\end{table}

		\newpage 
		
\begin{table}[H]
	\centering
	\caption{Factor Regressions for the Betting Against Beta factor sorts}
	\paragraph{}
	\begin{adjustbox}{scale=0.85, center}
		\begin{tabular}{|l|l|l|l|l|l|l|l|l|}
			\hline
			Portfolio sorts & xret & alpha & mkt & smb & hml & rmw & cma & umd \\ \hline
			~ & ~ & ~ & ~ & ~ & ~ & ~ & ~ & ~ \\ 
			1 (low beta) & 0.988 & 0.373 & 0.663 & 0.608 & 0.167 & 0.082 & 0.166 & -0.044 \\ 
			~ & [6.41] & [5.60] & [41.02] & [26.65] & [5.44] & [2.62] & [3.61] & [-2.79] \\ \hline
			~ & ~ & ~ & ~ & ~ & ~ & ~ & ~ & ~ \\ 
			2 & 0.995 & 0.216 & 0.959 & 0.797 & 0.149 & 0.068 & 0.111 & -0.095 \\  
			~ & [4.81] & [4.52] & [82.83] & [48.77] & [6.76] & [3.05] & [3.35] & [-8.39] \\ \hline
			~ & ~ & ~ & ~ & ~ & ~ & ~ & ~ & ~ \\ 
			3 (high beta) & 0.773 & 0.125 & 1.215 & 1.097 & 0.053 & -0.366 & -0.116 & -0.323 \\ 
			~ & [2.56] & [1.49] & [59.57] & [38.14] & [1.37] & [-9.27] & [-1.99] & [-16.25] \\ \hline
			~ & ~ & ~ & ~ & ~ & ~ & ~ & ~ & ~ \\ 
			BAB & 0.949 & 0.47 & 0.07 & 0.13 & 0.204 & 0.454 & 0.397 & 0.199 \\ 
			~ & [7.76] & [4.13] & [2.52] & [3.34] & [3.89] & [8.51] & [5.04] & [7.4] \\ \hline
			~ & ~ & ~ & ~ & ~ & ~ & ~ & ~ & ~ \\  
			BAB (net trading costs) & 0.851 & 0.372 & 0.069 & 0.131 & 0.204 & 0.452 & 0.397 & 0.197 \\ 
			~ & [6.97] & [3.28] & [2.5] & [3.37] & [3.9] & [8.5] & [5.06] & [7.36] \\ \hline
		\end{tabular}
	\end{adjustbox}
	\begin{tablenotes}
		\small
		\item \textbf{Notes:} The 5 Fama-French factors are: mkt (market risk factor), smb (Small minus Big), hml (High minus Low), rmw (Robust Minus Weak) and the cma (Conservative Minus Aggressive); to which we add umd (Momentum). The dependent variables are the equally-weighted tercile portfolio sorts on the universe of CRSP US stocks and the BAB factor with and without transaction costs. All variables are at monthly frequency with a sample from 1963:01-2021:12.
	\end{tablenotes}
\end{table}

\newpage 	
\begin{table}[!ht]
	\centering
	\caption{Factor Regressions for the Betting Against (Bad) Beta factor 3x3 sorts}
	\paragraph{}
	\begin{adjustbox}{scale=0.8, center}
		\begin{tabular}{|l|l|l|l|l|l|l|l|l|}
			\hline	
			3x3 portfolio sorts & xret & alpha & mkt & smb & hml & rmw & cma & umd \\ \hline
			~ & ~ & ~ & ~ & ~ & ~ & ~ & ~ & ~ \\ 
			1 (low beta/ low bad-beta) & 1.163 & 0.614 & 0.638 & 0.748 & 0.156 & -0.086 & 0.085 & -0.053 \\ 
			~ & [6.52] & [6.43] & [27.51] & [22.86] & [3.53] & [-1.93] & [1.28] & [-2.35] \\ \hline
			~ & ~ & ~ & ~ & ~ & ~ & ~ & ~ & ~ \\ 
			2 & 1.01 & 0.332 & 0.687 & 0.58 & 0.188 & 0.174 & 0.172 & -0.002 \\ 
			~ & [6.68] & [5.22] & [44.58] & [26.69] & [6.41] & [5.83] & [3.92] & [-0.11] \\ \hline
			~ & ~ & ~ & ~ & ~ & ~ & ~ & ~ & ~ \\ 
			3 & 0.916 & 0.194 & 0.755 & 0.63 & 0.074 & 0.156 & 0.323 & -0.032 \\ 
			~ & [5.28] & [2.16] & [34.62] & [20.47] & [1.79] & [3.69] & [5.19] & [-1.49] \\ \hline
			~ & ~ & ~ & ~ & ~ & ~ & ~ & ~ & ~ \\ 
			4 & 1.094 & 0.361 & 0.943 & 0.948 & 0.136 & -0.061 & 0.091 & -0.132 \\ 
			~ & [4.77] & [4.48] & [48.26] & [34.4] & [3.65] & [-1.62] & [1.63] & [-6.95] \\ \hline
			~ & ~ & ~ & ~ & ~ & ~ & ~ & ~ & ~ \\ 
			5 & 0.978 & 0.174 & 0.955 & 0.692 & 0.204 & 0.208 & 0.099 & -0.096 \\ 
			~ & [5.00] & [3.80] & [86.06] & [44.19] & [9.66] & [9.70] & [3.14] & [-8.86] \\ \hline
			~ & ~ & ~ & ~ & ~ & ~ & ~ & ~ & ~ \\ 
			6 & 0.938 & 0.136 & 0.988 & 0.754 & 0.105 & 0.062 & 0.152 & -0.061 \\ 
			~ & [4.49] & [2.27] & [68.16] & [36.9] & [3.82] & [2.20] & [3.68] & [-4.32] \\ \hline
			~ & ~ & ~ & ~ & ~ & ~ & ~ & ~ & ~ \\ 
			7 & 0.757 & 0.19 & 1.171 & 1.236 & 0.046 & -0.494 & -0.143 & -0.387 \\ 
			~ & [2.30] & [1.40] & [35.55] & [26.60] & [0.73] & [-7.76] & [-1.52] & [-12.04] \\ \hline
			~ & ~ & ~ & ~ & ~ & ~ & ~ & ~ & ~ \\ 
			8 & 0.828 & 0.098 & 1.206 & 0.962 & 0.187 & -0.169 & -0.076 & -0.298 \\ 
			~ & [2.97] & [1.30] & [66.01] & [37.32] & [5.37] & [-4.79] & [-1.46] & [-16.75] \\ \hline
			~ & ~ & ~ & ~ & ~ & ~ & ~ & ~ & ~ \\ 
			9 (high beta/ high bad-beta) & 0.738 & 0.054 & 1.222 & 1.055 & 0.021 & -0.394 & -0.043 & -0.263 \\
			~ & [2.49] & [0.64] & [58.77] & [35.95] & [0.53] & [-9.80] & [-0.72] & [-12.95] \\ \hline
			~ & ~ & ~ & ~ & ~ & ~ & ~ & ~ & ~ \\ 
			BABB & 1.253 & 0.919 & 0.048 & 0.359 & 0.178 & 0.186 & 0.233 & 0.117 \\ 
			~ & [8.38] & [6.08] & [1.31] & [6.94] & [2.56] & [2.62] & [2.22] & [3.27] \\ \hline
			~ & ~ & ~ & ~ & ~ & ~ & ~ & ~ & ~ \\ 
			BABB (net trading costs) & 1.003 & 0.674 & 0.045 & 0.354 & 0.184 & 0.182 & 0.224 & 0.115 \\
			~ & [6.73] & [4.47] & [1.24] & [6.86] & [2.65] & [2.57] & [2.15] & [3.23] \\ \hline
		\end{tabular}
	\end{adjustbox}
	
	\begin{tablenotes}
		\small
		\item \textbf{Notes:} The 5 Fama-French factors are: mkt (market risk factor), smb (Small minus Big), hml (High minus Low), rmw (Robust Minus Weak) and the cma (Conservative Minus Aggressive); to which we add umd (Momentum). The dependent variables are 3x3 equally-weighted portfolio sorts on the universe of CRSP US stocks, the BABB factor with and without transaction costs. All variables at monthly frequency with a sample from 1963:01-2021:12.
	\end{tablenotes}
	\label{t2}
\end{table}

	\end{document}